\documentclass{PoS}
\newcommand{\vcq}{V_{cd(s)}}
\newcommand{\DG}{\Delta\Gamma}
\newcommand{\ipb}{{\rm pb}^{-1}}
\newcommand{\ifb}{{\rm fb}^{-1}}

\newcommand{\bmath}{\begin{displaymath}}
\newcommand{\emath}{\end{displaymath}}

\newcommand{\fz}{f_+(0)}
\newcommand{\chargedkenu}{D^0\to K^- e^+ \nu_e}
\newcommand{\chargedpienu}{D^0\to \pi^- e^+ \nu_e}
\newcommand{\neutralkenu}{D^+ \to \bar{K}^0 e^+ \nu_e}
\newcommand{\neutralpienu}{D^+ \to \pi^0 e^+ \nu_e}

\newcommand{\dele}{\Delta E}
\newcommand{\mbc}{M_{{\rm BC}}}

\newcommand{\enu}{e^+ \nu_e}
\newcommand{\kk}{K^-K^+}

\newcommand{\DD}{D\bar{D}}

\newcommand{\vcs}{|V_{cs}|}
\newcommand{\vcd}{|V_{cd}|}

\newcommand{\beqn}{\begin{equation}}
\newcommand{\eeqn}{\end{equation}}

\newcommand{\ra}{\rightarrow}

\def\cleoc{\hbox{CLEO-c}}

\title{Recent Results on Charm Semileptonic Decays}

\ShortTitle{Charm Semileptonic Decays}

\author{\speaker{Bo Xin}%
         \\
        Purdue University\\
        E-mail: \email{bxin@purdue.edu}}

\abstract{We review the recent results on $D$ and $D_s$ meson semileptonic decays from \cleoc, {\it BABAR}, and Belle.
Comparisons with lattice quantum chromodynamics (LQCD) calculations and
implications for $B$ physics are also discussed.
}

\FullConference{Flavor Physics and CP Violation 2009\\
		 May 27-June 1 2009\\
		 Lake Placid, NY, USA }

\begin{document}

\section{Introduction}

In the Standard Model, the charge-changing transitions involving quarks
are described by the Cabibbo-Kobayashi-Maskawa~(CKM) matrix~\cite{ckm}.
Semileptonic decays are
the preferred way to determine the CKM matrix elements, 
because all the
strong interaction binding effects are confined to the hadronic
current, which can be parameterized by form factors.

However, the power of semileptonic decays in probing the CKM matrix has been
severely limited by our knowledge of the strong interaction effects.
While techniques such as lattice quantum chromodynamics (LQCD)~\cite{fnallqcd} offer increasingly precise calculations of the
hadronic form factors, experimental validation of these predictions is highly desired.
In charm semileptonic decays, the CKM matrix elements $\vcd$ and $\vcs$ are tightly constrained by CKM unitarity.
Therefore, precise measurements of charm semileptonic decay rates enable rigorous tests of
theoretical calculations of the form factors.
Because of the similarity between $D$ meson and $B$ meson semileptonic decays,
a validated theory can then be applied to the $B$ system with increased confidence.
In addition,
combining the measured $D$ meson semileptonic decay rates with
the theoretical calculations of form factors, such as those based on
Lattice QCD, direct measurments of the CKM matrix elements $\vcd$ and $\vcs$ can be made.
Studies of the exclusive semileptonic decays of the $D$ and $D_s$ mesons are also important to gain a complete understanding
of charm semileptonic decays, and 
as a probe of quark content and properties of the final state hadron.


\section{Experimental techniques}
\label{sec:tech}

In the last a few years, the experimental precision in charm semileptonic decays has been greatly improved.
Various experiments contributed to this improvement, including \cleoc, {\it BABAR}, and Belle~\footnote{FOCUS provides very precise measurements of $D$ lifetimes.
$D^0$ semileptonic decays are studied by FOCUS using the $D^0$ mesons from $D^*\ra D^0\pi^+$~\cite{pikenu_focus}.
These results are not covered in this review.
}.
The analysis techniques fall into two categories -- with tagging (tagged) and without tagging (untagged).

At \cleoc, both the tagged and untagged methods are used. We will focus on the tagged technique~\footnote{The
\cleoc ~untagged $D\ra K/\pi \enu$ results~\cite{281kpienuuntag} are superseded by the tagged results with a larger data sample.}.
The $D$ mesons are produced through the decays $e^+e^- \ra \psi(3770) \ra \DD$ at the center-of-mass energy near 3.770~GeV.
This is a particularly clean environment since there is not enough energy to produce any additional particles other than the $\DD$.
There are typically 5 -- 6 charged particles per event.
The presence of two $D$ mesons in a $\psi(3770)$ event allows
a tag sample to be defined in which a $\bar{D}$
is reconstructed in a hadronic decay mode.
A sub-sample is then defined in which a positron and a set of hadrons, as a signature
of a semileptonic decay, are required in addition to the tag.
Tagging a $\bar{D}$ meson in a $\psi(3770)$ decay provides a $D$
with known four-momentum, allowing a semileptonic decay to be
reconstructed with no kinematic ambiguity, even though the
neutrino is undetected.
Therefore, the reconstruction of the semileptonic side is almost background free.

\begin{figure*}[btb]
 \centering
  \includegraphics*[width=3.2in]{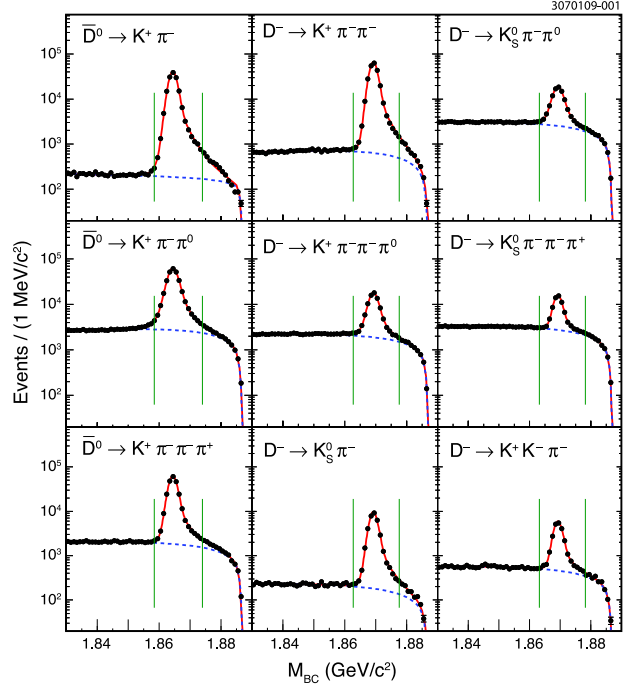}
  \caption{$M_{\rm BC}$ distributions using the 818 ${\rm pb}^{-1}$ $\psi\left(3770\right)\rightarrow D\bar{D}$ event sample from \cleoc.
  The background contributions are shown by the dotted lines.
  The vertical lines show the limits of the $M_{\rm BC}$ signal regions.
  }
  \label{fig:mbc_log}
\end{figure*}

Tagged events are selected based on two variables: $\dele \equiv
E_D - E_{{\rm beam}}$, the difference between the energy of the
$\bar{D}$ tag candidate ($E_D$) and the beam energy
($E_{{\rm beam}}$), and the beam-constrained mass $\mbc \equiv
\sqrt{E^2_{{\rm beam}}/c^4 - |\vec{p}_D|^2/c^2}$, where
$\vec{p}_D$ is the measured momentum of the $\bar{D}$ candidate.
The yield of each tag mode is obtained from a fit to the $\mbc$ distribution.
Using $818 \nolinebreak ~\mathrm{pb}^{-1}$ $\psi\left(3770\right)\rightarrow D\bar{D}$ event sample,
corresponding to approximately 5.4 million $D\bar{D}$ events,
\cleoc ~reconstructed approximately 660,000 neutral and 480,000 charged tags,
with an event tagging efficiency of about 20\%~\cite{818kpienu}.
The fits to the $\mbc$ distributions are shown in Fig.~\ref{fig:mbc_log}.

By performing an energy scan between 3.97 and 4.26 GeV, \cleoc ~determined the
center-of-mass energy that maximizes the yield of $D_s$ to be 4.170 GeV,
where the $D_s$ mesons are dominantly from $e^+e^-\ra D^*_s D_s$~\cite{dsscan}.
With a tagging technique similar to the one we described above, 
$D_s$ tag candidates are selected using hadronic final states.
They are then combined with well reconstructed photons to calculate the missing mass squared
$MM^{*2} = (E_{\rm CM}-E_{D^{\rm tag}_s}-E_{\gamma})^2-(\vec{p}_{\rm CM} - \vec{p}_{D^{\rm tag}_s}-\vec{p}_{\gamma} )^2$,
where $E_{\rm CM}$ ($\vec{p}_{\rm CM}$) is the center-of-mass energy (momentum),
$E_{D^{\rm tag}_s}$ ($\vec{p}_{D^{\rm tag}_s}$) is the tag energy (momentum),
and $E_\gamma$ ($\vec{p}_\gamma$) is the energy (momentum) of the additional photon.
Using the \cleoc ~600 ${\rm pb}^{-1}$ data set at 4.170 GeV,
corresponding to about 0.55 million $D_s^* D_s$ events,
about 44,000 good $D_s$ and photon combinations are found and used in
the reconstruction of the semileptonic $D_s$ decays.


Both {\it BABAR} and Belle study charm semileptonic decays using data collected at the $\Upsilon(4S)$.
The event environments are not as clean as the charm threshold. 
{\it BABAR} uses an untagged technique, where
the neutrino four-momentum is estimated from the other particles in the event.
The decays $D^0 \rightarrow K^- e^+ \nu_e (\gamma)$  are reconstructed in
$e^+e^- \rightarrow c\bar{c}$ events where the $D^0$ originates from
the $D^{*+} \rightarrow D^0 \pi^+$~\cite{babarkenu}.
At Belle, a novel technique has been used to analyze 
the events in which
$e^+e^-\rightarrow D_\mathrm{tag}^{(*)}D_\mathrm{sig}^{*-}X$,
$D_\mathrm{sig}^{*-}\rightarrow \bar D_\mathrm{sig}^0\pi^-_s$,
where $X$ may include additional $\pi^\pm$, $\pi^0$, or $K^\pm$ mesons,
and $\pi_s^-$ stands for a slow pion~\cite{bellekpilnu}.
By fully reconstructing the $D_\mathrm{tag}^{(*)}X\pi^-_s$,
the four-momentum of the $\bar D_\mathrm{sig}^0$ is known.
%

\section{Semileptonic decays of the $D$ mesons}

\subsection{$D$ semileptonic decays to $\pi$ and $K$ mesons from \cleoc, \emph{BABAR}, and Belle}

For pseudoscalar-to-pseudoscalar semileptonic decays, when the lepton mass is negligibly small,
the strong interaction dynamics can be described by a single form factor $f_+\left(q^2\right)$,
where $q^2$ is the invariant mass of the lepton-neutrino system.
The rate for a $D$ semileptonic decay
to a $\pi$ or $K$ meson
is given by
\begin{equation}
\frac{d\Gamma(D\rightarrow \pi(K) e\nu)}{dq^2}=X\frac{G_F^2\left|\vcq\right|^2}{24\pi^3}p^3\left|f_+\left(q^2\right)\right|^2,
\label{eq:diffrate}
\end{equation}
where $G_F$ is the Fermi constant, $\vcq$ is the relevant CKM matrix element,
$p$ is the momentum of the $\pi$ or $K$ meson in the rest frame of the parent $D$,
and $X$ is a multiplicative factor due to isospin, equal to 1 for all modes except $\neutralpienu$, where it is $1/2$.
Using the $818 \nolinebreak ~\mathrm{pb}^{-1}$ $\psi\left(3770\right)\rightarrow D\bar{D}$ event sample,
\cleoc ~measures the partial decay rates
$\DG=\int{\frac{d\Gamma}{dq^2}dq^2}$ in seven $q^2$ bins each for $\chargedpienu$ and $\neutralpienu$ and
nine $q^2$ bins each for $\chargedkenu$ and $\neutralkenu$.
The partial rates are then fit using several parameterizations of $f_+\left(q^2\right)$,
extracting form factor shape parameters, $\left|\vcq\right|\fz$, and branching fractions.
Taking estimates of $\fz$ from theory, $\vcd$ and $\vcs$ are also extracted~\cite{818kpienu}.

\begin{figure*}[bptb]
\centering
  \includegraphics*[width=3.7in]{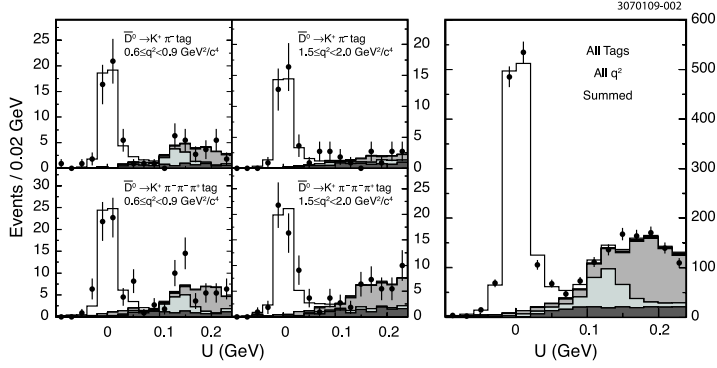}
  \caption{Fits of the $U$ distributions for $\chargedpienu$ from \cleoc ~(only a subset are shown).
  The unshaded histograms are signal.
  See Ref.~\cite{818kpienu} for details of the background components.}
  \label{fig:u_pi}
\end{figure*}

After a tag is identified, a positron and a set of hadrons are searched for in the recoiling system against the tag.
Semileptonic decays are identified using the variable $U \equiv
E_{\rm miss} - c|\vec{p}_{\rm miss}|$, where $E_{\rm miss}$ and
$\vec{p}_{\rm miss}$ are the missing energy and momentum of the
$D$ meson decaying semileptonically, calculated using the
difference of the four-momentum of the tag and that of the observed
products of the semileptonic decay.
Signal yields are extracted from $U$ distributions.
Properly reconstructed decays are separated from backgrounds using an unbinned maximum-likelihood fit,
executed independently for each semileptonic mode, each tag mode, and each $q^2$ bin.
A sample of the $U$ distributions for $\chargedpienu$ is shown in Fig.~\ref{fig:u_pi}.
The signal and background shapes of the fits are taken from Monte Carlo samples.

The partial rates are then obtained by inverting the efficiency matrices, which account for both efficiency
and the smearing across $q^2$ bins.
The $q^2$ resolutions
averaged over the entire $q^2$ range are about
0.008~(GeV$/c^2)^2$ for $D^0 \to \pi^- e^+ \nu_e$, $D^0 \to K^-
e^+ \nu_e$ and  $D^+ \to \bar{K}^0 e^+ \nu_e$, and approximately
0.014~(GeV$/c^2)^2$ for $D^+ \to \pi^0 e^+ \nu_e$.
Least squares fits are made to these partial rates, using several form factor parameterizations.
Short surveys of these form factor parameterizations can be found in Refs.~\cite{818kpienu} and \cite{281kpienu} and references therein.
As the data do not support the physical basis of the simple pole~\cite{richmanpole}, modified pole~\cite{modpole}, and ISGW2~\cite{ISGW2} parameterizations,
the model-independent series expansion~\cite{srp} is generally of most interest.

\begin{figure}[!tb]
\begin{center}
\includegraphics[height=1.9in]{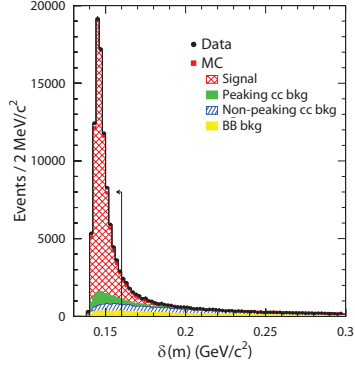}
\caption{Comparison of the $\delta(m)$ distributions from the {\it BABAR} data and simulated events.
The arrow indicates the additional selection applied
for the $q^2$ distribution measurement.}
\label{fig:deltamback}
\end{center}
\end{figure}

Using 75 ${\rm fb}^{-1}$ of data collected at $\Upsilon(4S)$,
{\it BABAR} studies $\chargedkenu$ where the $D^0$ originates from
the $D^{*+} \rightarrow D^0 \pi^+$~\cite{babarkenu}.
An untagged technique, as described in Section~\ref{sec:tech}, is used.
Each $D^0$ candidate is then combined with a charged pion.
The mass difference
$\delta(m) = m(D^0 \pi^+)-m(D^0)$ is evaluated and is shown in Fig.~\ref{fig:deltamback}.
The $q^2$ resolution is about 20 times larger than the \cleoc ~measurement.
Note that {\it BABAR} measures the yield of $\chargedkenu$ relative to the reference decay channel $D^0\ra K^-\pi^+$,
and uses ${\mathcal B}(D^0\ra K^-\pi^+)$ from PDG~\cite{PDG2006} for normalization.

The Belle measurement of $D^0\ra \pi l \nu$ and $K l\nu$ branching fractions and form factors is performed
using 282 $\ifb$ of data at the $\Upsilon(4S)$. Here $l$ denotes both electrons and muons.
The full event reconstruction method greatly improves the resolutions of the missing mass squared and the $q^2$ distributions.
The $q^2$ resolution is about 2 times larger than the \cleoc ~measurement.


A comparison of the recent branching fraction measurements for $\chargedpienu$, $\chargedkenu$, $\neutralpienu$, and $\neutralkenu$ is shown in Table~\ref{tab:bf_comp}.
Good agreement is seen between the results obtained by different experiments.
The \cleoc ~818 $\ipb$ results are the most precise, with a 1.4\% uncertainty on ${\mathcal B}(\chargedkenu)$
and 3.0\% uncertainty on ${\mathcal B}(\chargedpienu)$.

Form factor shapes are revealed by removing the kinematic terms from the right-hand-side
 of Eq.~\ref{eq:diffrate} and assuming the unitarity values for $\vcd$ and $\vcs$. 
In Fig.~\ref{fig:kpienuvslqcd}, the \cleoc ~results 
are compared between isospin conjugate modes and with the latest LQCD calculations~\cite{fnallqcd}.
The \cleoc ~results agree with LQCD calculations within uncertainties,
but are much more precise.
The LQCD bands are obtained using the modified pole model~\cite{modpole}.
In Table~\ref{tab:alpha_comp},
we compare the form factor shape parameter $\alpha$ as measured by different experiments with LQCD calculations.
The \cleoc ~results are obtained by simultaneously fitting the isospin-conjugate modes.
The semileptonic decays of the $D^+$ mesons have not been studied at the $B$ factories.
The experimental measurements of $\alpha$ are generally compatible with each other.
However, determining the level of agreement between experiments and LQCD is difficult,
because of their sensitivities to different $q^2$ regions,
and the uncertainties due to the use of the modified pole model.

\begin{table}[!tbp]
\begin{center}
{\small
\begin{tabular}{lcccc}\hline\hline
 & $\chargedpienu$ & $\chargedkenu$ & $\neutralpienu$ & $\neutralkenu$ \\ \hline
Belle  (282 $\ifb$)~\cite{bellekpilnu}          & 0.279(27)(16) & 3.45(10)(19)      &               &              \\
{\it BABAR}  (75 $\ifb$)~\cite{babarkenu}           &               & 3.522(27)(45)(65) &               &              \\
CLEO-c (818 $\ipb$)~\cite{818kpienu}                 & 0.288(8)(3)   & 3.50(3)(4)        & 0.405(16)(9)  & 8.83(10)(20) \\ \hline\hline
\end{tabular}
}
\caption{Comparison of branching fraction results (\%) from different experiments.}
\label{tab:bf_comp}
\end{center}
\end{table}


\begin{figure}[bptb]
\centering
  \includegraphics*[width=2.4in]{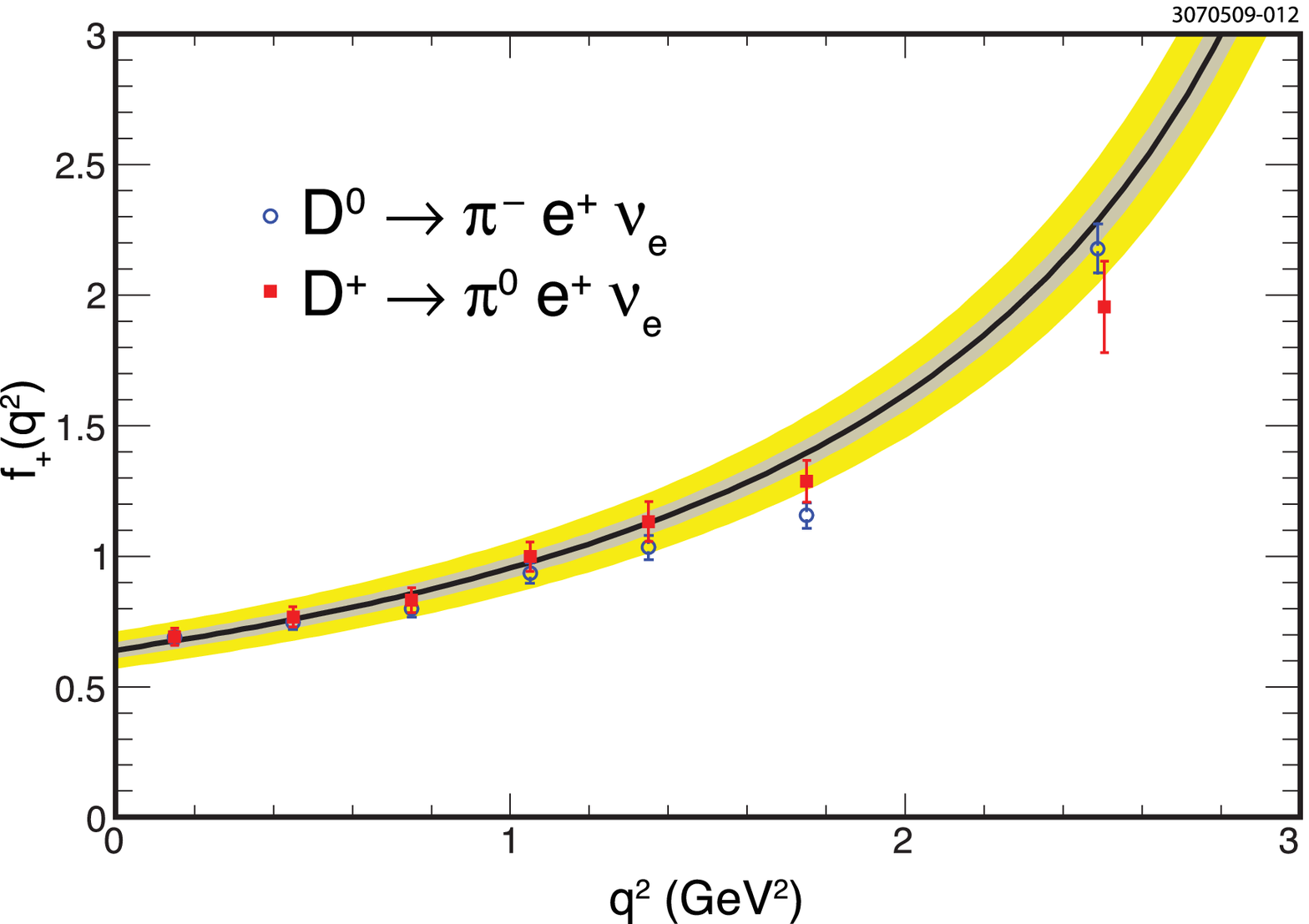}
  \includegraphics*[width=2.4in]{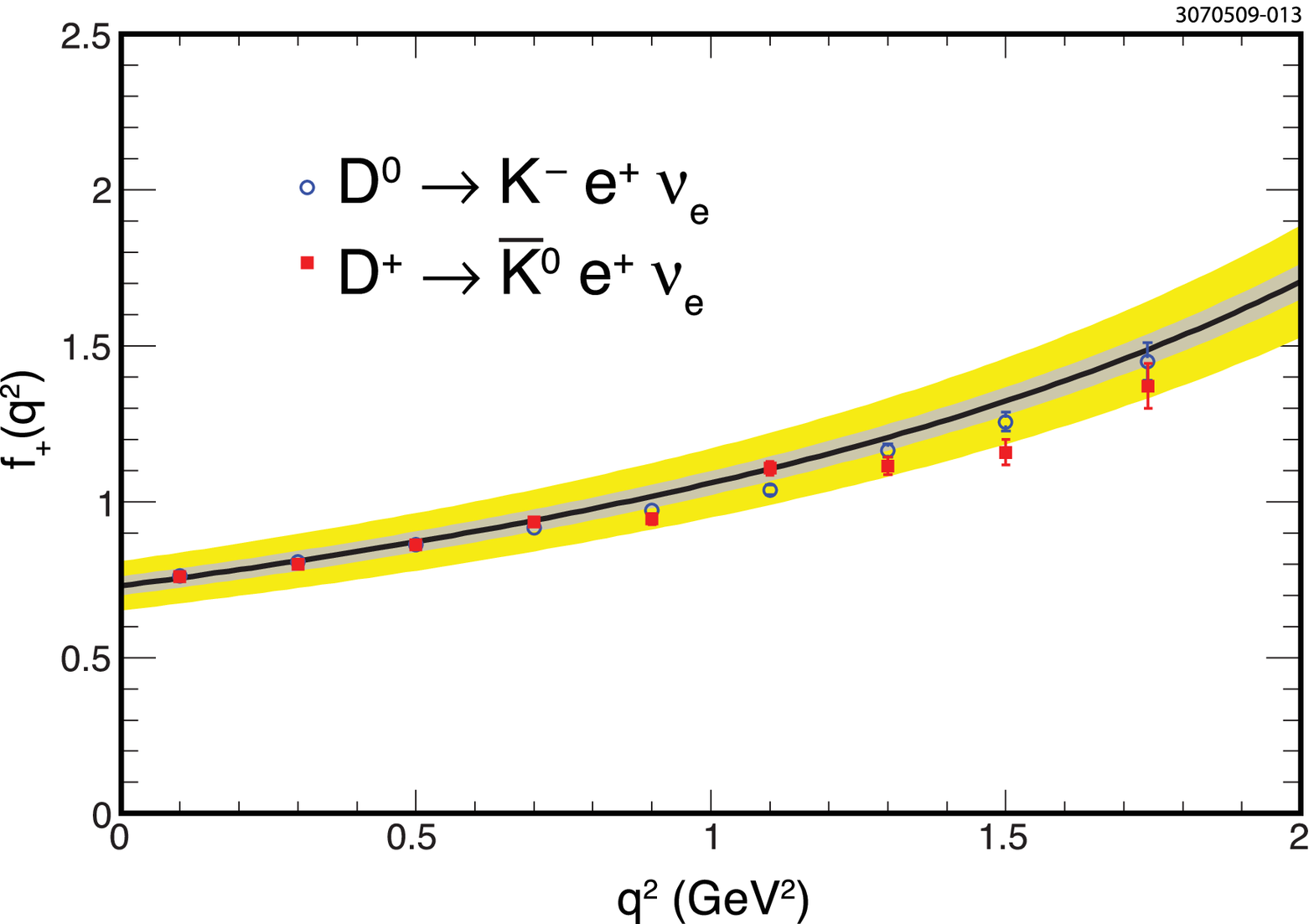}
  \caption{$f_+(q^2)$ comparison between isospin conjugate modes from \cleoc
  ~and with LQCD calculations~\cite{fnallqcd}.
    The solid lines represent LQCD fits to the modified pole model. 
    The inner bands show LQCD statistical uncertainties, and the outer bands the sum in quadrature of LQCD statistical and systematic uncertainties.}
  \label{fig:kpienuvslqcd}
\end{figure}

\begin{table}
\begin{center}
{\small
\begin{tabular}{lcc}
\hline\hline
 & $\alpha^K$  & $\alpha^\pi$ \rule[-1mm]{-1mm}{4.3mm}\\ \hline
LQCD \cite{fnallqcd} & 0.50(4)(7) & 0.44(4)(7) \rule[-1mm]{-1mm}{4.3mm} \\
Belle \cite{bellekpilnu} & 0.52(8)(6) & 0.10(21)(10) \\
{\it BABAR} \cite{babarkenu} & 0.377(23)(29) & \\
CLEO-c (818 $\ipb$) \cite{818kpienu}  & 0.30(3)(1) & 0.21(7)(2) \\
\hline\hline
\end{tabular}
}
\end{center}
\caption{Comparison of form factor shape parameter $\alpha$, from fits using the modified pole model.}
\label{tab:alpha_comp}
\end{table}

Taking the $\left|\vcq\right|\fz$ values from the isospin-combined three parameter series expansion fits and using the LQCD measurements
for $f_+(0)$, \cleoc ~
finds
$\vcd = 0.234\pm0.007\pm0.002\pm0.025$ and $\vcs = 0.985\pm0.009\pm0.006\pm0.103$,
where the third uncertainties are from the LQCD calculation of $f_+(0)$.  These are in agreement with those based on the assumption of CKM unitarity
$\vcs = 0.97334 \pm 0.00023$ and $\vcd = 0.2256 \pm 0.0010$~\cite{PDG2008}.
The \cleoc ~$\vcs$ measurement is the most precise direct determination. The $\vcd$ measurement is the most precise using semileptonic decays.


%

\subsection{$D$ semileptonic decays to vector mesons from \cleoc}

Using one-third of its full data sample, \cleoc ~has studied the form factors in $P\ra V$ transitions.
Among these studies, the form factor in $D\ra\rho\enu$ is of special interest~\cite{281rhoenu}.
When combined with the form factor in $D\ra K^* \enu$, the form factor in $D\ra\rho\enu$ helps in determining $|V_{\rm ub}|$
using the double ratio method~\cite{rhoenudoubleratio}.
In addition, this is the first form factor measurement in Cabibbo suppressed $P\ra V$ transitions.
\cleoc ~finds ${\mathcal B}(D^0\ra\rho^-\enu) = (1.56\pm0.16\pm0.09)\times 10^{-3}$
and ${\mathcal B}(D^+\ra\rho^0\enu) = (2.32\pm0.20\pm0.12)\times 10^{-3}$.
A four-dimensional log likelihood fit is performed to the isospin-conjugate modes simultaneously,
the form factor ratios~\cite{richmanpole} are found to be
$R_V = 1.40 \pm 0.25 \pm 0.03$ and $R_2 = 0.57 \pm 0.18 \pm 0.06$.

\subsection{Observations of new semileptonic modes from \cleoc}

In addition to studying the existing modes with unprecedented precision, \cleoc ~also has many results
from its searches for new semileptonic modes. These modes include
$D^+\ra \eta\enu$~\cite{281etaenu}, $D^0\ra \rho^-\enu$, $D^+ \ra \omega \enu$~\cite{281rhoenu},
and $D^0\ra K^-\pi^+\pi^-\enu$~\cite{kpipienu}.
These new modes are important for gaining a complete understanding of charm semileptonic decays.

\section{$D_s$ semileptonic decays}

\subsection{$D_s$ exclusive semileptonic decays from \cleoc}

The first absolute branching fraction measurements of the $D_s$ semileptonic decays
have been made by \cleoc~\cite{dsexclu} using 310 $\ipb$ of data at 4.170 GeV.
Via the tagged analysis technique,
six exclusive semileptonic modes
are searched for. The missing-mass-squared ($MM^2$) distributions are shown in Fig.~\ref{fig:mm2}.
Among these, ${\mathcal B}(D_s^+\to K^0 e^+ \nu_e)=(0.37\pm 0.10 \pm 0.02)$\% and ${\cal B}(D_s^+\to K^{*0} e^+
\nu_e)=(0.18 \pm 0.07 \pm 0.01)$\% are the first measurements of Cabibbo suppressed exclusive $D_s$ semileptonic
decays. The measurement of
${\cal B}(D_s^+\to f_0 e^+ \nu_e) \times {\cal B}(f_0
\to \pi^+\pi^-) =(0.13\pm 0.04 \pm 0.01)$\% is the first
direct evidence of a semileptonic decay including a scalar meson
in the final state.

By searching for several additional hadronic final states formed with two charge tracks with or without a $\pi^0$,
\cleoc ~finds no evidence for other $D_s$ semileptonic decays.
However, the total width of these measured exclusive modes is about 16\% lower than the $D^0$ and $D^+$ semileptonic widths,
Theoretical interpretations include
SU(3) symmetry breaking and possibly non-factorizable contributions.
The measured ratio ${\mathcal B}(D_s^+\to \eta^\prime e^+ \nu_e)/{\mathcal B}(D_s^+\to \eta e^+ \nu_e) = 0.36 \pm 0.14$
also sheds light on $\eta - \eta^\prime$ glueball mixing~\cite{dsexclu}.

\begin{figure}[!bt]
\begin{center}
\mbox{
\includegraphics[width=2.3in]{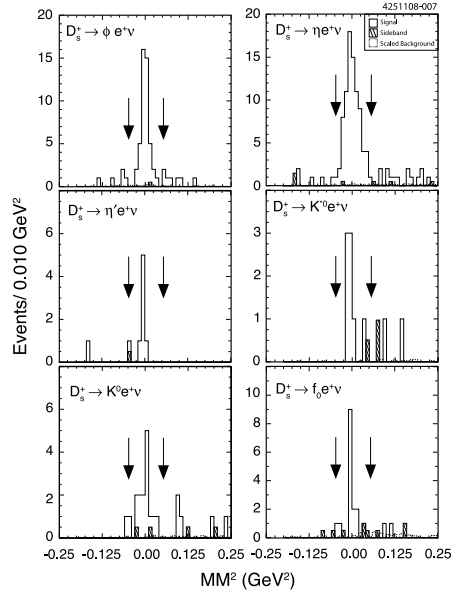}
}
\caption{The $\rm MM^2$ distribution for tagged $D_s$ semileptonic
events in the exclusive modes from the \cleoc ~310 $\ipb$ of data.
See Ref.~\cite{dsexclu} for details on background components.
} \label{fig:mm2}
\end{center}
\end{figure}

\subsection{$D_s^+ \ra\kk\enu$ from \emph{BABAR}}

{\it BABAR} has also studied the decay $D_s^+\ra \kk\enu$~\cite{babarkkenu}. In this mode, because of the higher mass of the spectator $s$-quark,
LQCD calculates the form factors more accurately.
{\it BABAR} uses the same experimental method as used in its $D^0 \ra K^-\enu$ analysis, except that here no $D^*$ is used.
This measurement is normalized to the branching fraction of $D_s^+\ra\kk\pi^+$ measured by \cleoc.
Babar finds ${\cal B}(D_s^+\ra \phi \enu) = (2.61\pm0.03\pm0.08\pm0.15)$\%, where the last uncertainty is due to
${\cal B}(D_s^+\ra\kk\pi^+)$,
and finds a small S-wave contribution, possibly $f_0 \ra \kk$, corresponding to ($0.22^{+0.12}_{-0.08}\pm0.03$)\% of the $\kk\enu$ decay rate.

\begin{figure}[!tbp]
\begin{center}
\mbox{
\includegraphics[width=2.7in]{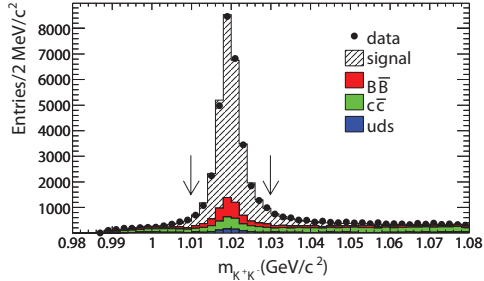}
}
\caption{The $K^+K^-$ invariant mass distribution from {\it BABAR} with 214 $\ifb$ of data and simulated events.
The arrows indicate the selected $K^+K^-$ mass interval.}
\label{fig:mkk}
\end{center}
\end{figure}

\subsection{$D_s^+\ra f_0(980)\enu$ from \cleoc}

$D_s$ semileptonic decays provide a very clean environment to study the properties of the $f_0(980)$ meson.
Using 600 $\ipb$ of data at 4.170 GeV, \cleoc ~studies the decay $D_s^+\ra f_0(980)\enu$~\cite{f0enu}.
There is another important motivation for this study --
it is suggested that the decay $B_s\ra J/\psi f_0(980)$ can be an alternative to $B_s \ra J/\psi \phi$ in measuring
the CP Violation effects in the $B_s$ system~\cite{stonejpsif0}.

The analysis is performed using the \cleoc ~tagging technique.
The $\pi^+\pi^-$ and $\kk$ invariant mass
distributions are shown in Fig.~\ref{f0mass}.
The branching fractions
${\cal B}(D_s^+\to f_0(980) e^+ \nu)\times {\cal B}(f_0\to \pi^+\pi^-)=(0.20\pm 0.03\pm 0.01)$\% and ${\cal B}(D_s^+\to \phi e^+ \nu)=(2.36\pm 0.23\pm 0.13)\%$
are measured with improved precision.

The dependence of the decay rates on $q^2$ has been investigated.
At $q^2$ of zero, the ratio
$\left[\frac{d{\cal B}}{dq^2}(D_s^+\to f_0 e^+ \nu) {\cal B}(f_0\to \pi^+\pi^-)\right]/
\left[\frac{d{\cal B}}{dq^2}(D_s^+\to \phi e^+ \nu) {\cal B}(\phi\to K^+K^-)\right]$,
which has been predicted to equal
$\left[{\cal B}(B_s\to J/\psi f_0){\cal B}(f_0\to \pi^+\pi^-)\right]/\left[{\cal B}(B_s\to J/\psi \phi){\cal B}(\phi\to K^+K^-)\right]$,
is measured to be ($42\pm11$)\%, indicating that $B_s \ra J/\psi f_0$ would be a very useful place to study
CP violation in the $B_s$ system.
We note that $J/\psi f_0$ is a CP eigenstate, and so no angular analysis is needed.

By fitting the $\pi^+\pi^-$ invariant mass spectrum using a relativistic Breit-Wigner function,
the mass and width of the $f_0(980)$ are determined to be
$(977^{+11}_{-9}\pm 1){\rm~MeV}$ and $(91^{+30}_{-22}\pm 3)~{\rm MeV}$, respectively.
The partial rates are fit to the simple pole model, the pole mass is found to be
$(1.7^{+4.5}_{-0.7}\pm 0.2$)~GeV.

\begin{figure}[!tb]
\centering
  \includegraphics*[width=4in]{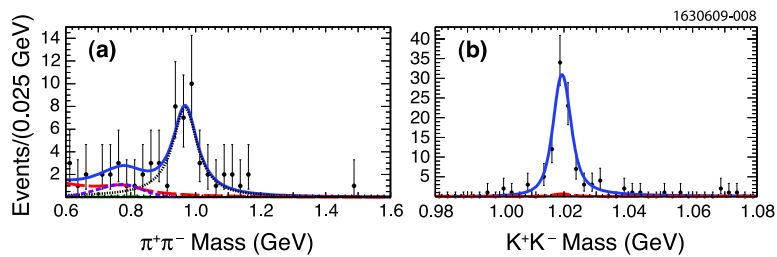}
 \caption{The invariant mass distribution for (a) $\pi^+\pi^-$ and (b) the $K^+ K^-$ in the semileptonic mode from \cleoc ~with 600 $\ipb$ of data.
See Ref.~\cite{f0enu} for details on background components.}
 \label{f0mass}
 \end{figure}

\section{Summary and prospects}

The great progresses made by LQCD in the past a few years need validations.
Charm semileptonic decays have come to meet the challenges, serving as an excellent testing ground of LQCD.
Great improvements in the experimental precision have been achieved,
primarily due to the ever larger luminosities of data accumulated by the $B$-factories,
and more importantly, the clean event environment and powerful analysis technique employed by the \cleoc ~experiment.

Looking into the future, more precise LQCD calculations of charm semileptonic form factors are expected later this year.
More exciting results from the above mentioned experiments are on the way.
Novel event reconstruction techniques are being tried.
Many results are to be updated using larger data samples.
The larger data samples will enable some measurements that were previously impossible.
In the longer term future, BESIII will take the field into a new era of precision measurements.

\acknowledgments

I thank the organizers for such a wonderful conference in Lake Placid.
Valuable discussions with C. Davies are appreciated.
I. Shipsey is thanked for very helpful discussions and suggestions on this manuscript.

\end{document}